\documentclass[twocolumn]{aastex631}

\newcommand{\hMpc}{{\ifmmode{\ h^{-1}{\rm Mpc}}\else{$h^{-1}$Mpc\ }\fi}}  
\newcommand{\hGpc}{{\ifmmode{\ h^{-1}{\rm Gpc}}\else{$h^{-1}$Gpc\ }\fi}}  
\newcommand{\hmpc}{{\ifmmode{\ h^{-1}{\rm Mpc}}\else{$h^{-1}$Mpc\ }\fi}}  
\newcommand{\hkpc}{{\ifmmode{\ h^{-1}{\rm kpc}}\else{$h^{-1}$kpc\ }\fi}}  
\newcommand{\hMsun}{{\ifmmode{\ h^{-1}{\rm {M_{\odot}}}}\else{$h^{-1}{\rm{M_{\odot}}}$}\fi}}  
\newcommand{\kms}{{\ifmmode{\ {\rm km\, {s}^{-1}}}\else{\ km s$^{-1}$ }\fi}}  
\newcommand{\Msun}{{\ifmmode{\ {\rm {M_{\odot}}}}\else{${\rm{M_{\odot}}}$ }\fi}}  

\shorttitle{Convergence of the MW and M31 kinematics}
\shortauthors{Forero-Romero \& Sierra-Porta}
\graphicspath{{./}{figures/}}

\begin{document}

\title{On the Convergence of the Milky Way and M31 Kinematics from Cosmological Simulations}

\correspondingauthor{J. E. Forero-Romero}
\email{je.forero@uniandes.edu.co}

\author[0000-0002-2890-3725]{J. E. Forero-Romero}
\affiliation{Departamento de F\'isica \\
Universidad de los Andes\\
Cra. 1 No. 18A-10, Edificio Ip\\
  CP 111711\ Bogot\'a, Colombia}
  
  \affiliation{Observatorio Astron\'omico \\
Universidad de los Andes\\
Cra. 1 No. 18A-10, Edificio H\\
  CP 111711\ Bogot\'a, Colombia}

\author[0000-0003-3461-1347]{D. Sierra-Porta}
\affiliation{Departamento de F\'isica \\
Universidad de los Andes\\
Cra. 1 No. 18A-10, Edificio Ip\\
  CP 111711\ Bogot\'a, Colombia}
\affiliation{Facultad de Ciencias Básicas \\ Universidad Tecnológica de Bolivar \\ Cartagena de Indias 130010, Colombia.}



\begin{abstract}
The kinematics of the Milky Way (MW) and M31, the dominant galaxies in the Local Group (LG), can be used to estimate the LG total mass.
New results on the M31 proper motion have recently been used to improve that estimate.
Those results are based on kinematic priors that are sometimes guided and evaluated using cosmological N-body simulations.
However, the kinematic properties of simulated LG analogues could be biased due to the effective power spectrum truncation induced by the small size of the parent simulation.
Here we explore the dependence of LG kinematics on the simulation box size to argue that  cosmological simulations need a box size on the order of 1 Gpc in order to claim convergence on the LG kinematic properties.
Using a large enough simulation, we find  M31 tangential and radial velocities relative to the MW to be in the range $v_{\mathrm {tan}}=105^{+94}_{-59} \kms$ and $v_{\mathrm {rad}}=-108^{+68}_{-81}\kms$, respectively.
This study highlights that LG kinematics derived from N-body simulations have to be carefully interpreted taking into account the size of the parent simulation.
\end{abstract}



\section{Introduction}
\label{sec:intro}

The kinematics of the dominant galaxies in the Local Group (LG), the Milky Way (MW) and M31, can be used to estimate its total mass using a general approach known as the Timing Argument (TA) \citep{1959ApJ...130..705K,1982MNRAS.199...67E}. 
Two key measurements to be used as an input in the TA are the tangential and radial velocities of M31 relative to the MW.
The new results on the M31 proper motion (PM) provided by the Gaia satellite have spurred new activity to measure the tangential speed of M31 relative to the MW and therefore improve the LG total mass constraints \citep{2019ApJ...872...24V,2021MNRAS.507.2592S}.

However, translating PM measurements into a relative tangential velocity requires defining a prior on the tangential speed $v_{\rm tan}$.
Different priors lead to different PM results.
For instance, \cite{2008ApJ...678..187V} and \cite{2019ApJ...872...24V} take a flat prior on $v_{\rm tan}$  while \cite{2021MNRAS.507.2592S} take a prior proportional to $v_{\rm tan}$.
These two different shapes for the prior distribution result on different expectation values for the tangential speed, which can be interpreted as  \cite{2019ApJ...872...24V} having a preference towards lower tangential speeds than \cite{2021MNRAS.507.2592S}.

Cosmological N-body simulation in the Lambda Cold Dark Matter (LCDM) paradigm have been used to inform those priors \citep{2008ApJ...678..187V} and  to calibrate possible biases in the TA \citep{2008MNRAS.384.1459L}.
Simulations have also provided a numerically derived prior to place observed LG kinematics in a cosmological context \citep{2011MNRAS.417.1434F} and to constrain the total LG mass  independently of the analytical expressions derived from the TA \citep{gonzalez14}.

In general, the results from numerical simulations produce priors that favors values close to $80 \kms$ for the  tangential speed \citep{2011MNRAS.417.1434F,CLUES,APOSTLE,Sawala16,2016MNRAS.460L...5C} with some zoom simulations even favoring values with a median around $50 \kms$ \citep{ELVIS,HESTIA}

In this manuscript we show that these results might not be robust as a consequence of a relatively small simulation box size.
Here we argue that reaching convergence on the kinematic properties requires simulations on the order of 1 Gpc in box size, which to this date represents a challenge for numerical models of galaxy formation in a explicit cosmological context.

As a probe of kinematic convergence we use the LG barycenter speed.
The barycenter speed offers three advantages: it has a robust theoretical prediction as a function of the simulation box size, it is easy to measure in simulations and has a low uncertainty measurement for our LG. 
We use the theoretical baseline for the barycenter speed as an independent verification of the trends we find in simulations.
Finally, we use a cosmological simulation with box size close to 3 Gpc to report the preferred ranges for the radial and tangential speed of M31 with respect to the MW.

This article is structured as follows.
We start in Section \ref{sec:simulations} by describing the cosmological simulations we use to measure the kinematics for LG analogues.
We continue in Section \ref{sec:samples} with the detailed description of how we define a LG analogue in simulations.
We move into Section \ref{sec:predictions} to review the expectations from linear theory for the probability density function for the barycenter speed.
In Section \ref{sec:results} we present and discuss our results.
We close with our conclusions in Section \ref{sec:conclusions}.

\section{Cosmological Simulations}
\label{sec:simulations}

We use simulations from two different projects: IllustrisTNG \citep{2018MNRAS.475..624N,2018MNRAS.480.5113M,2018MNRAS.475..648P,2018MNRAS.475..676S,2018MNRAS.477.1206N,2019MNRAS.490.3196P,2019MNRAS.490.3234N} and  AbacusSummit \citep{Garrison2018,Garrison2019,Garrison2021,Maksimova2021,Hadzhiyska2021}.
They have different numerical setups and methods to find dark matter halos, although
their cosmological parameters are similar.
In the following we describe the most relevant features from each project.

\subsection{IllustrisTNG}

The IllustrisTNG project is a set of gravito-magnetohydrodynamical simulations with models for the physics of galaxy formation and evolution through cosmic time.
The simulations couple dark matter, cosmic gas, luminous stars and supermassive black holes in a redshift range from $z=127$ to the $z=0$.

The TNG simulations were performed on three different cubic volumes of different size and at three different resolutions.
This gives us a total of nine different simulations that we use in this work.
The simulations for each box size receive different names: TNG100 and TNG300, where the number indicates the box size of $106.5$ and $302.6$ in units of Mpc, respectively.
All these  simulations use the same cosmological parameters from the of the Planck 2015 results \citep{2016A&A...594A..11P} with a present time Hubble parameter of $H=67.74$ \kms Mpc$^{-1}$, present time dark energy density $\Omega_{\Lambda}=0.6911$, present time matter density $\Omega_m=0.3089$, power spectrum normalization $\sigma_8=0.8159$ and spectral index $n_s=0.9667$.
Table \ref{tab:MB_scale} lists the particle mass resolution for all the simulations.

As a proxy for a MW/M31 galaxy we use the main substructure inside the dark matter halo detected with the Friend-of-Friends (FOF) algorithm. 
We use masses defined by spheres that enclose $\Delta_c$ times the critical density of the Universe, where $\Delta_c$ is derived from the fitting formula in \cite{1998ApJ...495...80B}.
However, we use the maximum circular velocity as a selection criterion which has a weak dependence on different overdensity criteria to define the halo boundary \citep{2011ApJ...740..102K}.
Our results use the snapshot at a redshift of $z=0.1$ in order to allow a comparison against the results from AbacusSummit.

\subsection{AbacusSummit}

AbacusSummit is a suite of large dark matter only simulations. 
From this project we use a total of seven different simulations.

Five of the boxes were generated with the same global cosmological and numerical parameters, but only differ on the initial seed for the initial conditions. 
They have a box size of $L_{\rm box}=2967$ Mpc on a side. 
In that volume the dark matter distribution was sampled with $6912^3$ particles, which corresponds to a particle mass of $2\times 10^{9} \hMsun$.
We refer to these boxes as AbacusBase.

Two more boxes have a box size of $L_{\rm box}=1483$ Mpc on a side, one of them sampled with $6300^3$ particles (about $6\times 10^{8} \hMsun$ per particle) and the other with $3456^3$ particles ($2\times 10^{9} \hMsun$ per particle). 
We refer to these two boxes as AbacusHigh and AbacusHighBase, respectively.

The cosmological parameters on all these simulations follow the Planck 2018 cosmology \citep{Planck2018} with a present time Hubble parameter of $H=67.36$ \kms Mpc$^{-1}$, present time dark energy density $\Omega_{\Lambda}=0.685$, present time matter density $\Omega_m=0.315$, power spectrum normalization $\sigma_8=0.811$ and spectral index $n_s=0.96$. 

For our analysis we use the halo catalogs built on the snapshot at redshift of $z=0.1$ using the CompaSO algorithm \citep{Hadzhiyska2021}.
We use as a selection criterion the maximum circular velocity computed on the dominant substructure inside the Level1 halo defined by CompaSO. 
We use the masses for Level1 halos, which correspond to the same definition we use for IllustrisTNG halos.

\section{Local Group Analogue Definitions}
\label{sec:samples}

We aim at finding pairs of dark matter halos that broadly resemble the LG's mass and isolation.
We follow similar conditions as used by \cite{2011MNRAS.417.1434F}.
We start by selecting all halos with maximum circular velocities $V_{\rm max}\geq 200\kms$.
Then, we use these halos to find what we call an isolated pair.
Isolated pairs are two halos, $A$ and $B$, that are mutually their nearest halo.
We use a convention where $A$ refers to the least massive halo in the pair.
Furthermore, halos $A$ and $B$ do not have any other third halo more massive than halo $B$ closer than three times the pair separation.  
Finally, we only keep pairs were both halos have maximum circular velocities in the range $200 \kms \leq V_{\rm max}\leq 260\kms$, separations less than 1.5 Mpc and negative radial velocity after the Hubble-Lema\^itre expansion term is taken into account.

These selection criteria do not exclude the possibility that a group or cluster halo could be found near the pair, which could represent a significant perturbation not present in the observed LG. 
To estimate the impact of such configurations we look for pairs with a halo within five times the pair separation with $V_{\rm max}$ greater than $300\kms$. We find that this situation affects less than $1\%$ of the pairs. This presents negligible consequences for the statistical results presented in the paper.

\section{Kinematics from Linear Theory}
\label{sec:predictions}

\cite{ShethDiaferio} used linear theory extrapolated from Gaussian initial conditions to explicitly show that (if the ranges of halo masses and local background density are narrow) 
the velocity components for a halo population should follow a normal distribution.
As a consequence, the peculiar speed should follow a Maxwell-Boltzmann (MB) distribution.

Although \cite{ShethDiaferio} did not  consider the case of halo pairs, we argue that the barycenter velocity, being the sum of normally distributed variables (i.e. the velocity components of each pair member), will also have normally distributed components, which translates into an MB distribution for the barycenter speed.
We show in the next Section that it is indeed the case.

The normalized probability density function (PDF) for $v_{b}$ can thus be the written as
\begin{equation}
    P(v_{b})=\sqrt{\frac{2}{\pi}}\frac{v_{b}^2e^{-v_{b}^2/(2\sigma_b^2)}}{\sigma_b^3},
\end{equation}
where the scale $\sigma_b$ is a parameter with velocity dimensions that uniquely determines the distribution.

The cumulative distribution function (CDF) can then be written as
\begin{equation}
    P(<v_{b})=\mathrm{erf}\left(\frac{1}{\sqrt{2}}\frac{v_{b}}{\sigma_b}\right)-\sqrt{\frac{2}{\pi}}\frac{v_{b}e^{-v_{b}^2/(2\sigma_b^2)}}{\sigma_b},
    \label{eq:cumulative}
\end{equation}
where, $\mathrm{erf}(x)$ is the error function.


The linear extrapolation by \cite{ShethDiaferio} provides the following expression to compute $\sigma_b$ for halos of mass $m$

\begin{equation}
    \sigma_b(m) =  H_{0} \Omega_m^{0.6} \sigma_{-1}\sqrt{1-\frac{\sigma_0^4}{\sigma_1^2\sigma_{-1}^2}},
\label{eq:linear}
\end{equation}
where $H_0$ is the Hubble parameter at present time, $\Omega_m$ is the matter density parameter at present time and $\sigma_j$ are moment integrals of the matter power spectrum, $P(k)$,
\begin{equation}
    \sigma_j^2 (m) = \frac{1}{2\pi^2}\int_0^\infty dk k^{2 + 2j}P(k) 
    W^{2}(kR(m)),
\end{equation}
where $W(x)$ is the Fourier transform of the window function, and $R(m)$ is the virial radius associated with a halo of mass $m$ at a given redshift $z$.

Having a finite box size in a simulations means that the power spectrum is effectively truncated to $P(k)=0$ for $k<2\pi/L_{\mathrm {box}}$.
In the next section we show that this truncation adequately reproduces the barycenter speed trends as a function of the simulation box size.

Here we use a Top-Hat filter in real space, for which $W(x)= (3/x^3)(\sin(x) -x\cos(x))$.
For the typical halo size, $R(m)$, we pick a value of $0.25$ Mpc.
Considering larger different values for $R(m)$ does not have a significant impact on the results.  
We use the linearly extrapolated power spectrum down to $z=0.1$ with the analytical transfer function by \cite{EisensteinHu} and Planck 2015 cosmological parameters.

\section{Results and Discussion}
\label{sec:results}

\begin{figure}[ht!]
  \includegraphics[width=1.0\columnwidth]{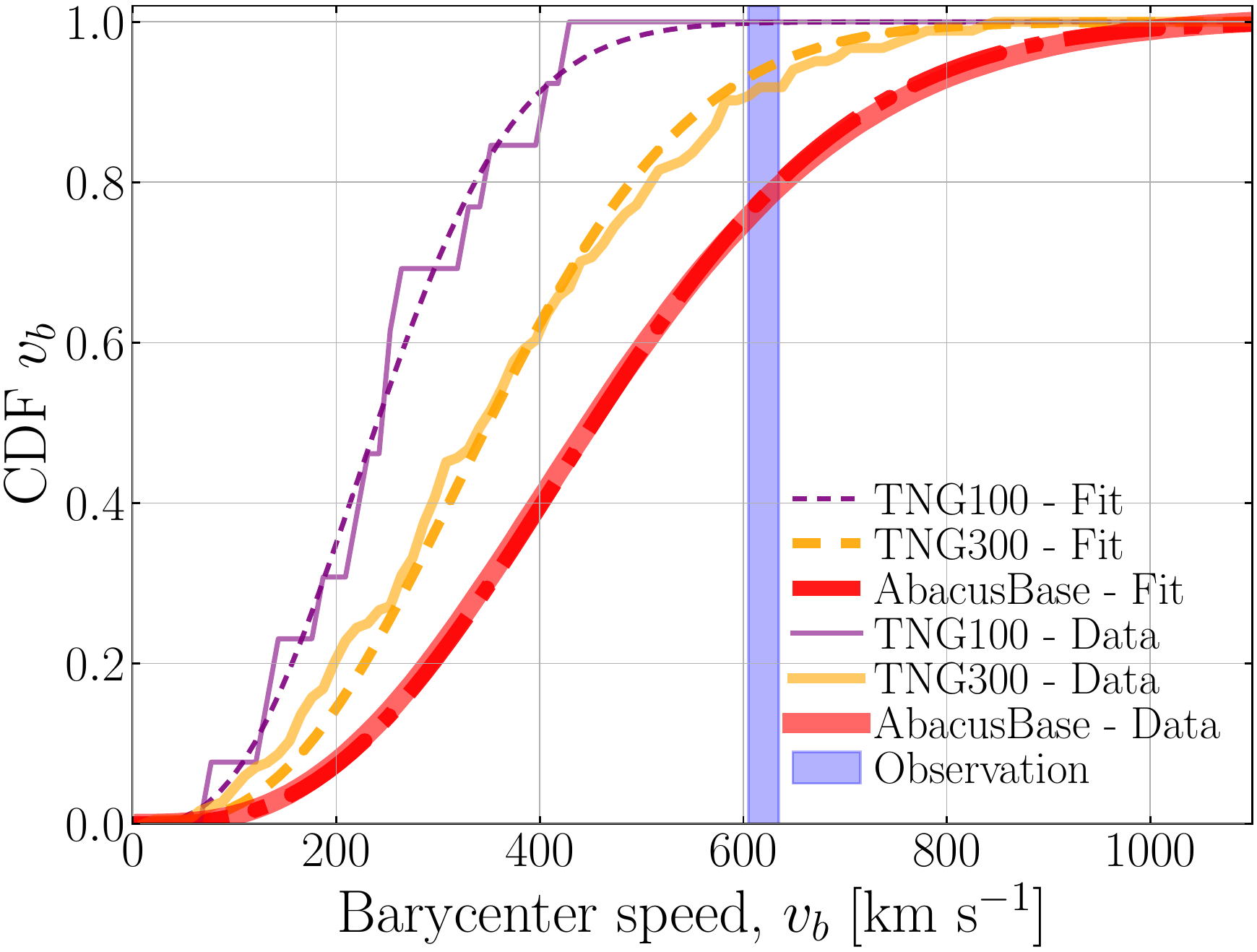}
\caption{Cumulative distribution for the barycenter speed from different cosmological simulations (continuous line) together with its best Maxwell-Boltzmann fit (dashed line). Larger box sizes correspond to distributions skewed towards higher speeds. The vertical stripe indicates the observational value for our LG.}
\label{fig:cumulativas}
\end{figure} 

Figure \ref{fig:cumulativas} shows the barycenter speed CDF
computed from different simulations together with its best MB fit, 
showing that the CDF for $v_b$ is well approximated by a MB CDF, as expected. 
As a reference value we plot the observed barycenter speed for our LG . 
This value is well determined by the dipole anisotropy in the Cosmic Microwave Background (CMB) \citep{FixenDipole} with a value of $620\pm 15 \kms$ derived from a careful review of the dynamics of the relative movement of the Sun respect to the LG and data from the Planck satellite \citep{Planck2018}.
For the largest simulations in our sample we estimate that the percentage of LG analogues with a barycenter speed equal or larger than the observed value is $p_{LG}=(21\pm3)\%$.

Figure \ref{fig:cumulativas} clearly shows that the peculiar velocities in an N-body cosmological simulation have a strong dependendence on the parent box size. 
Larger box sizes allow the development of DM halos with larger peculiar speeds.
Using simulations with small box sizes could give the erroneous impression of LG analogues with small peculiar speeds that could not reproduce the observed value for our LG.
What would be the box size beyond which the barycenter speed distribution is expected to converge?

To answer this question, first we estimate $\sigma_b$ as a function of the box size for all the simulation we have available.
Then we compare those numerical results against the expectations from linear theory to argue that for box sizes of at least $1$ Gpc one should expect the desired convergence.

\begin{table*}
    \centering
    \begin{tabular}{ccccc}
    \hline
    Simulation name & $\sigma_b$  [\kms]& $\Delta_{\sigma_b}$  [\kms]& $m_p$ [$10^{7} \Msun$] & $N_p$\\
    \hline
TNG100\_1 & 155 & 17 & 0.7 & 13 \\
TNG100\_2 & 122 & 28 & 6 & 6 \\
TNG100\_3 & 165  & 21 & 48 & 6 \\
TNG300\_1 & 226 & 7 & 6 & 184\\
TNG300\_2 & 220 & 8 & 47  & 133 \\
TNG300\_3 & 246 & 16 & 380 & 67\\
AbacusHigh & 295.1 & 0.8 & 90 & 32896\\
AbacusHighBase & 296.2 & 0.7 & 300 & 33936\\
AbacusBase & 298.9 & 0.3 & 300 & 269543\\
    \hline
    \end{tabular}
    \caption{Summary of all results for the characteristic speed in the Maxwell-Boltzmann distribution for the different simulations explored in this paper. 
    First column, simulation identifier. Second column, the characteristic speed. Third column, the uncertainty on the characteristic speed. 
    Fourth column, the mass of a single computational dark matter particle in the simulation.
    Fifth column, the number of LG analogues found in the simulation.
    The results for AbacusBase correspond to the mean value over the five simulations.}
    \label{tab:MB_scale}
\end{table*}

\begin{figure}[ht!]
  \includegraphics[width=1.0\columnwidth]{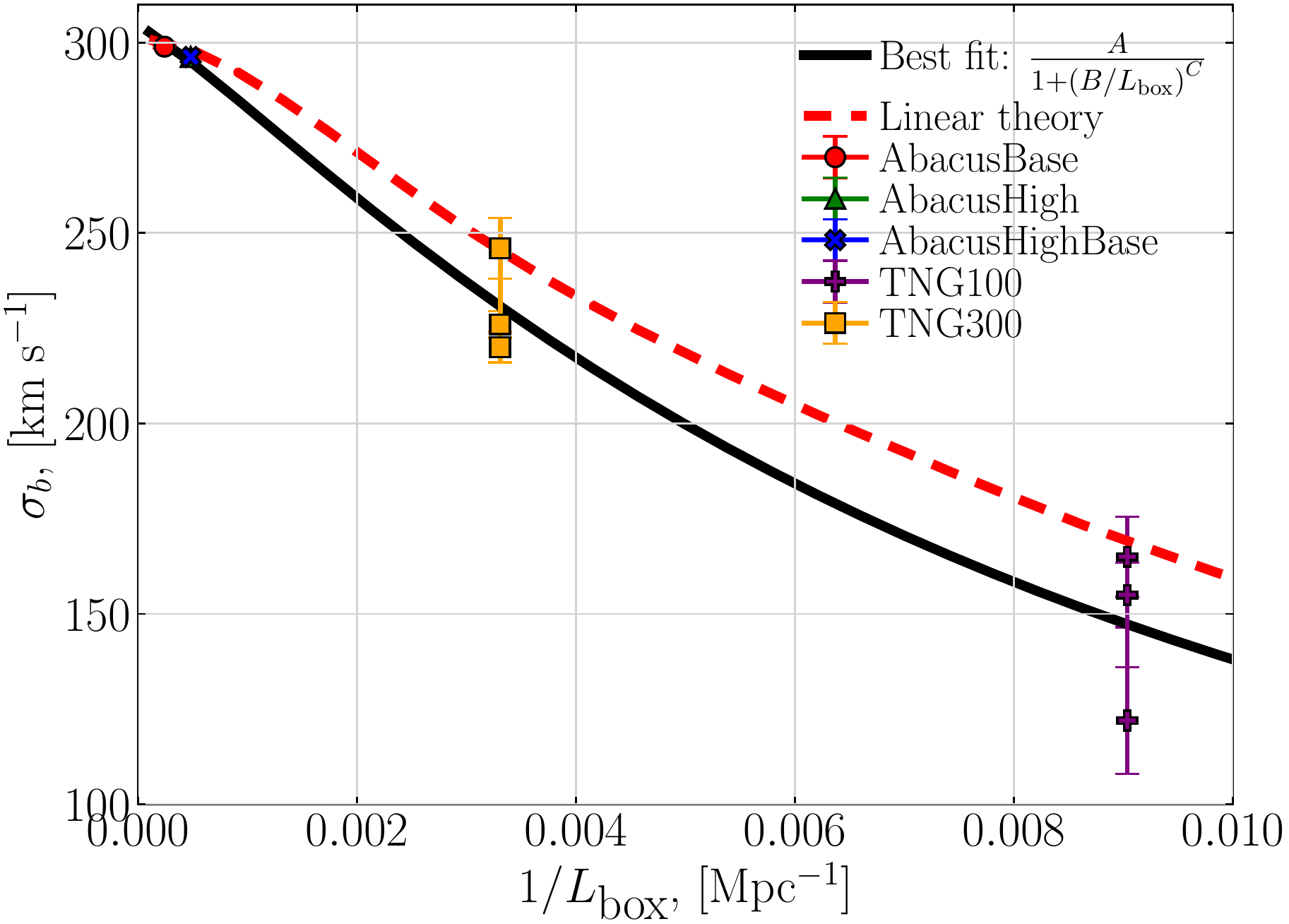}
\caption{Dependence of the speed scale $\sigma_b$ as a function of the inverse of the simulation box size.
The symbols present the results from simulations. 
The dashed line corresponds to the linear theory predictions described in Eq. (\ref{eq:linear}), in this case the box size indicates the wavelength cut below which the power spectrum is suppressed.
The continuous line corresponds the function in Eq. (\ref{eq:fit}) with the parameters that best fit the the simulation results.}
\label{fig:fit}
\end{figure}

Table \ref{tab:MB_scale} presents our $\sigma_b$ estimates for the simulations we have at hand 
where we observe that the dominant influence on $\sigma_b$ comes from the simulation box size.
This is more evident in Figure \ref{fig:fit}, where we compare $\sigma_b$ as a function of the inverse box size both from the simulations and the linear theory expectations (dashed line) after imposing a power spectrum truncation to mimic the effect of a finite box size.
To compute the results from Eq. (\ref{eq:linear}) we use $10^4$ Mpc as the largest box size.
Considering larger box sizes does not have a significant impact on those results.

\begin{figure*}[]
  \includegraphics[width=1.0\columnwidth]{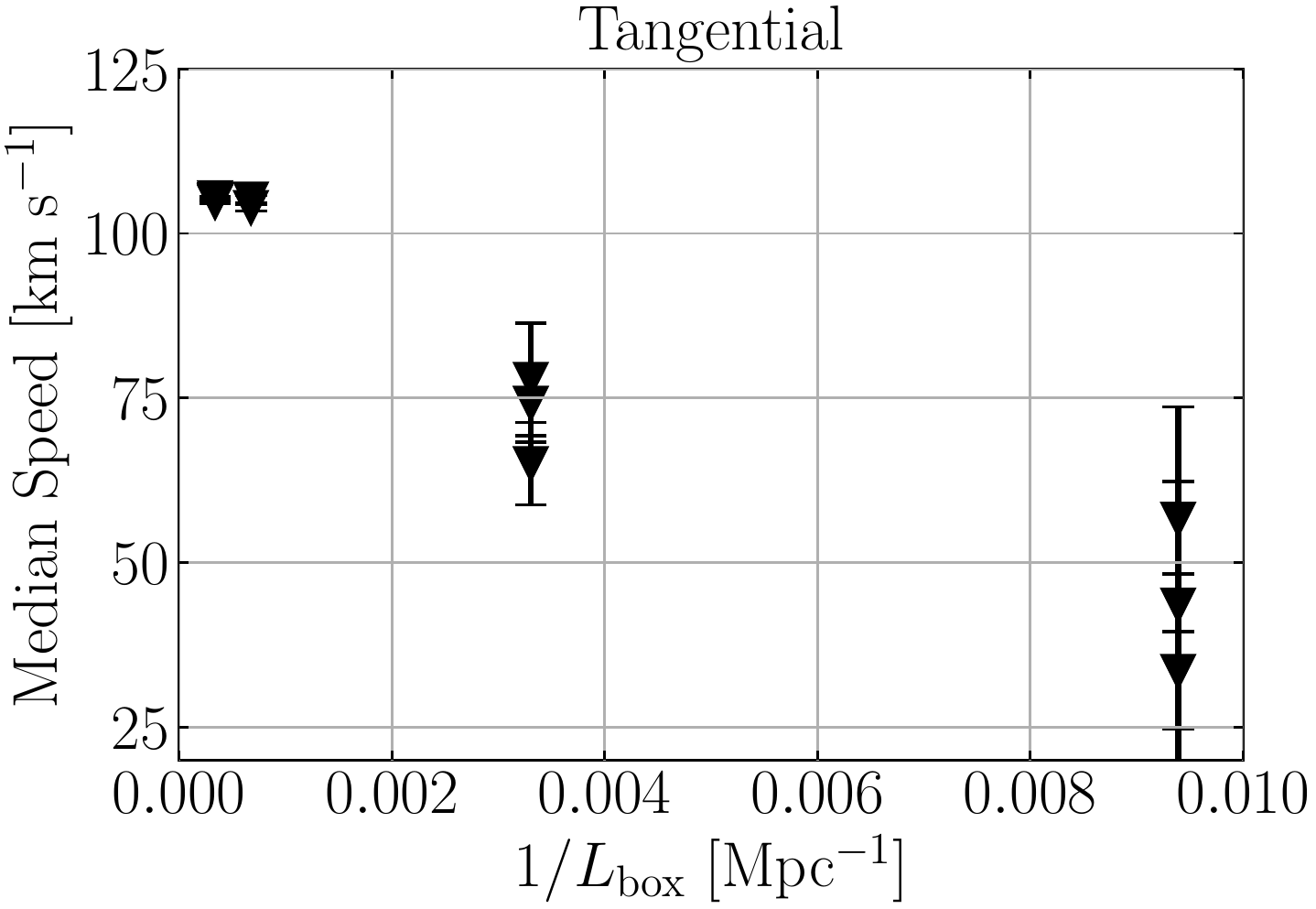}
\includegraphics[width=1.0\columnwidth]{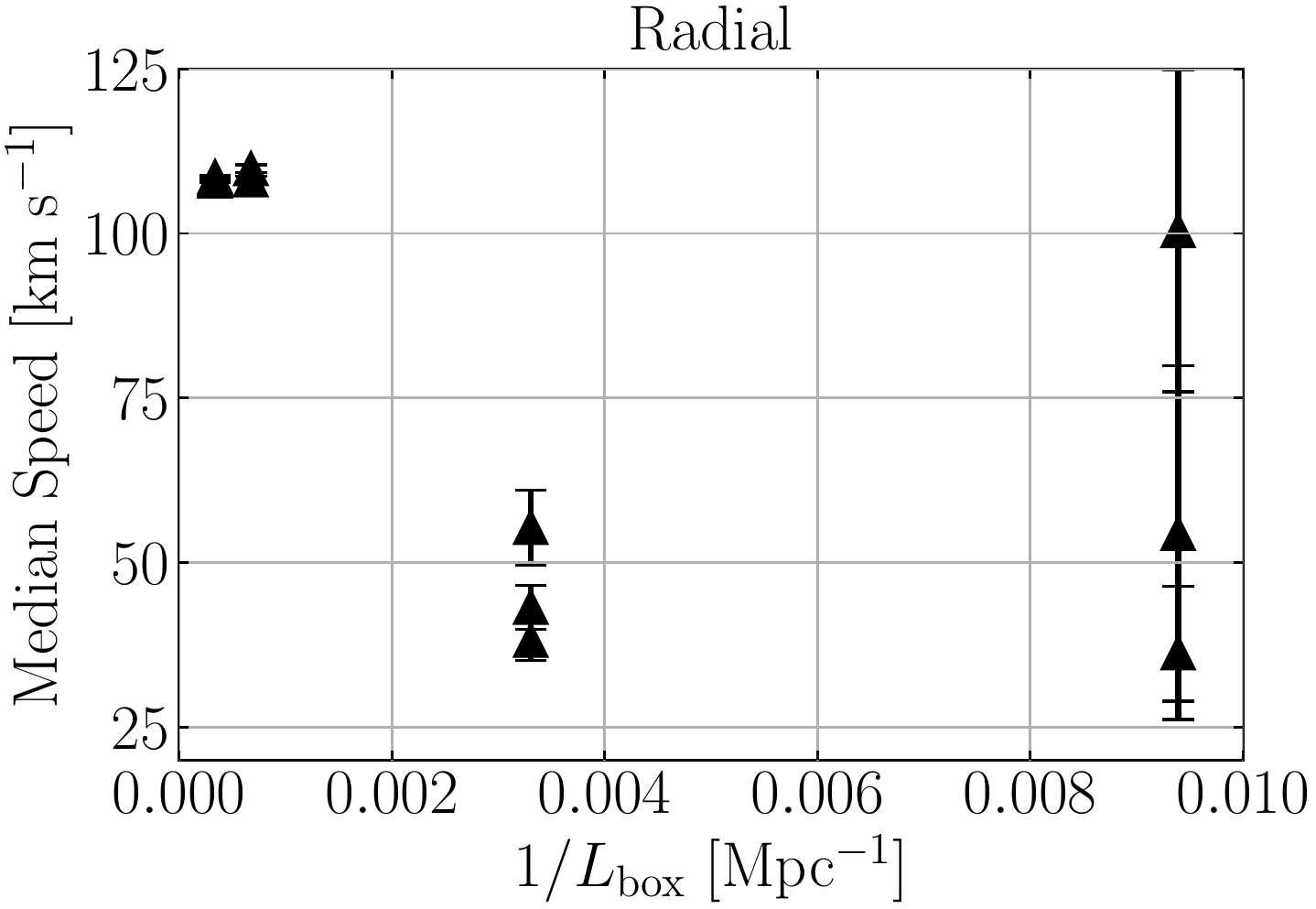}
\caption{Median tangential (left) and radial (right) speeds of M31 relative to the MW as a function of the inverse simulation box size. All the simulations listed in Table \ref{tab:MB_scale} are included in this Figure.}
\label{fig:median}
\end{figure*} 

\begin{figure}[]
  \includegraphics[width=1.0\columnwidth]{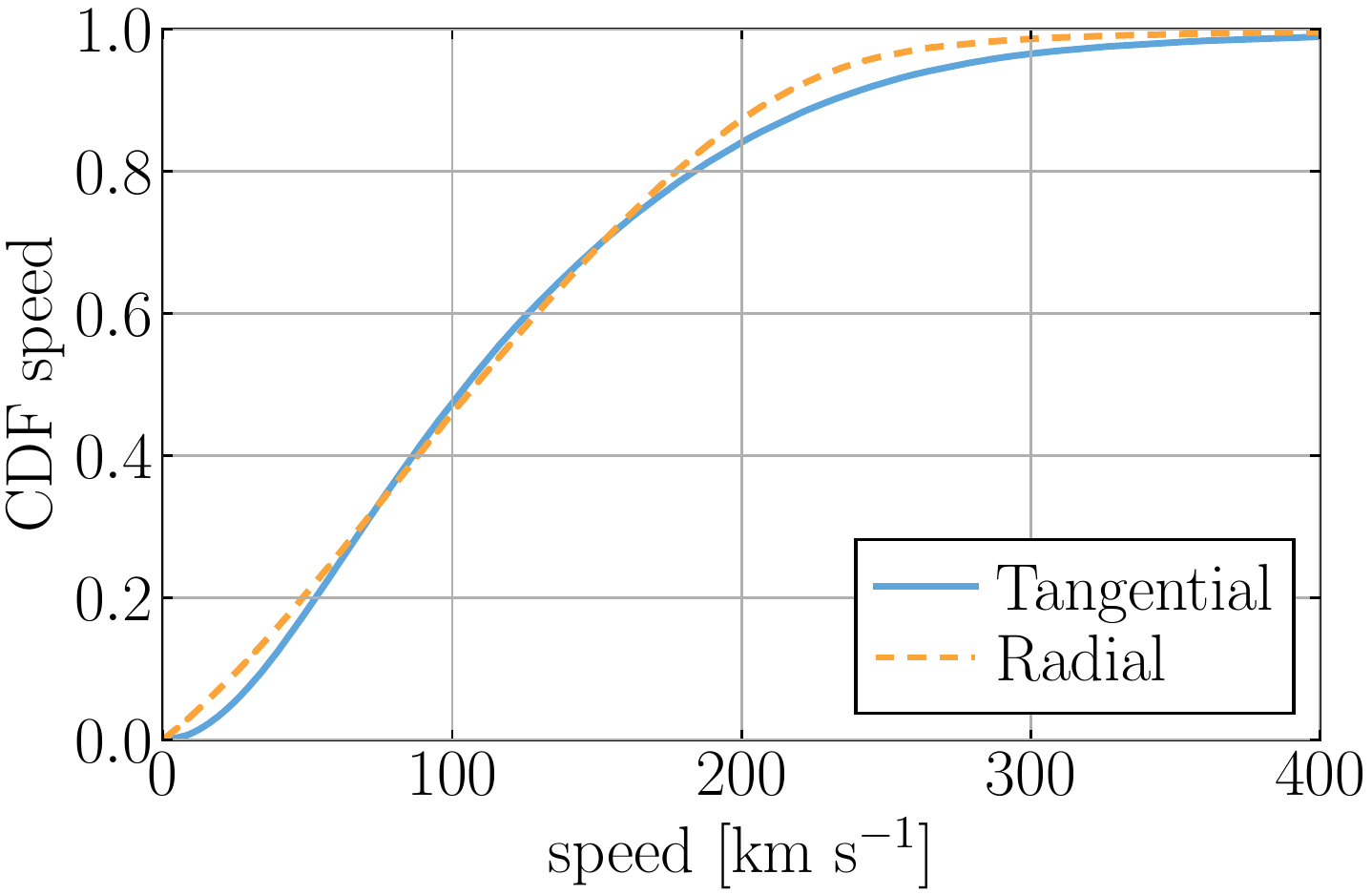}
\caption{Cumulative distribution for the tangential and radial speeds of M31 relative to the MW. Only pairs from AbacusBase (the simulation with the largest box size) are included in this Figure.}
\label{fig:dependence}
\end{figure} 

We observe that linear theory successfully accounts for the overall $\sigma_b$ dependence on $L_{\mathrm{box}}$, although it slightly overestimates the results from N-body simulations as already shown by \cite{ShethDiaferio}.
The value that we obtain from linear theory for $\sigma_b$ in the limit of infinite box size is \textbf{$300 \kms$}.

Figure \ref{fig:fit} also shows the best least-squares fit to the simulation data. 
We use the function
\begin{equation}
\sigma_b=\frac{A}{1+(Bx)^C},
\label{eq:fit}
\end{equation}
with $x=1/L_{\mathrm{box}}$ and $A>0$, $B>0$, $C>1$.
The advantages of this functional form are that in the limit of $x$ towards to zero, $\sigma_b$ tends to a finite value with a null derivative.

The best fit has  $A=304\pm 12 \kms$, $B=117\pm 12$ Mpc and $C=1.2\pm 0.2$. 
This yields  $\sigma_b=304 \pm 12 \kms$ in the limit of infinite box size, consistent with the value derived from the AbacusBase simulations ($298.9 \pm 0.3 \kms$) and linear theory ($300\kms$) showing that a converged estimate of the peculiar speed distributions from simulations requires a box size on the order of $1$ Gpc.

So far, these results establish that: (a) the peculiar velocity distribution strongly depends on the parent box size and (b) one could expect convergence of the barycenter kinematics for box sizes on the order of $1$ Gpc.
However, the question of what is the box size influence on the tangential and radial M31 speed relative to the MW, remains open.

We address that question in Figure \ref{fig:median}.
Using all the available simulations, we compute
the median in the radial and tangential speed distributions.
We estimate the uncertainty from 1000 bootstrapping iterations.
There, we find a strong dependence of the median on the box size. 
These results show a maximum median speed for the box sizes above $1$ Gpc and suggest a minimum for the smallest box sizes, with the caveat of having large error bars for the smallest box sizes due to the low number of LG analogues found in those volumes.

Finally, we use all pairs from the AbacusBase simulations to present in Figure \ref{fig:dependence} the full CDF for the tangential and radial speed. 
These volumes are large enough to provide a robust estimate 
for those CDFs. 
Above box sizes of $1$ Gpc we do not expect strong fluctuations for the peculiar speed distributions that are in turn used to measure the relative tangential and radial speeds.

From these results we find the tangential and radial M31 velocities relative to the MW to be in the range  $v_{\rm tan}=105^{+94}_{-59} \kms$ and $v_{\rm rad}=-108^{+68}_{-81} \kms$, where the central value corresponds to the median and the uncertainties are computed to match the $16$-th and $84$-th percentiles, with negative velocities standing for infalling motion.

These results support the statement that tangential velocity estimates from  simulations with box sizes on the order of $100$ Mpc might underestimate the median of the true prior distribution expected from LCDM. 
This includes zoom simulations of constrained realizations with a parent N-body simulation built to reproduce the observed large scale structure around the LG. \citep{2011MNRAS.417.1434F,CLUES,APOSTLE,Sawala16,2016MNRAS.460L...5C}.

For instance, three LG pairs from constrained simulations, in the CLUES project, with a box size of $87$ Mpc on a side have radial and tangential speeds lower than $70$ \kms and $50$\kms, respectively \citep{2013ApJ...767L...5F}. 
Twelve LG pairs from the ELVIS project that correspond to zoom simulations from a parent cosmological box of $70.4$ Mpc show median values for the radial and tangential speed of $57$ \kms and $43$ \kms, respectively \citep{ELVIS}. 
The thirteen intermediate resolution LG pairs from the HESTIA project that come from constrained realizations simulated on a box of 147.5 Mpc on a side \citep{HESTIA} present median radial and tangential speeds of $61.7$ \kms and $48.3$ \kms, respectively. 
All these values are consistently lower than our estimates from simulations with converged kinematics.

\section{Conclusions}
\label{sec:conclusions}

In this paper we presented a study of simulated LG kinematics from cosmological N-body simulations as a function of the simulation box size.
Combining the results from different simulations we showed that there is a strong dependence of the LG barycenter speed as a function of the simulation box size.
Larger box sizes correspond to wider speed distributions.

We use linear theory to show that this trend can be understood in terms of the power spectrum truncation due to a finite box size.
Using these results from simulations and linear theory we find that converged results for the barycenter speed can be expected for box sizes on the order of $1$ Gpc and above.

We also study the changes in the tangential and radial velocity of M31 relative to the MW as a function of the box size.
There we also find a strong dependence whereby the largest box sizes correspond to the larger tangential and radial speeds.
From the simulations with the largest box size ($3$ Gpc) in our sample  we estimate the tangential and radial M31 velocities relative to the MW to be in the range  $v_{\rm tan}=105^{+94}_{-59} \kms$ and $v_{\rm rad}=-108^{+68}_{-81} \kms$.

These findings suggest that LG kinematics derived from cosmological simulations with box sizes on the order of a few $100$ Mpc  might favor low tangential speed values as a consequence of a relatively small box size.
In that event, one has to be cautious in the comparison of the tangential velocities in simulations against observations, keeping in mind the influence of the simulation box size.

Numerical studies performed to understand the LG formation and evolution in a cosmological context, that want to claim convergence on the LG kinematics, will have to tackle the computational challenge imposed by a box size that must be on the order of $1$ Gpc,  while having enough resolution to correctly describe scales on the order of $100$ kpc.
Recent results based on a constrained simulation with $1$ Gpc box size on a side, performed to study the LG, are the first ones to go in that direction \citep{2022MNRAS.512.5823M}.

\bibliography{ms}{}
\bibliographystyle{aasjournal}

\end{document}